\documentclass[graybox,natbib,nosecnum]{svmult}
\bibpunct{(}{)}{;}{a}{}{,} 

\pdfoutput=1   

\usepackage{mathptmx}       
\usepackage{helvet}         
\usepackage{courier}        
\usepackage{type1cm}        

\usepackage{makeidx}         
\usepackage{graphicx}        
\usepackage{multicol}        
\usepackage[bottom]{footmisc}
\usepackage[normalem]{ulem}	
\usepackage{hyperref}  

\usepackage{soul}   

\newcommand{\ms}{\mbox{m s$^{-1}$}}
\newcommand{\kms}{\mbox{km s$^{-1}$}}

\newcommand{\teff}{$T_{\rm eff}~$}

\makeindex             

\begin{document}

\title*{55 Cancri (Copernicus): A Multi-Planet System with a Hot Super-Earth and a Jupiter Analogue}
\author{Debra A. Fischer}
\institute{Debra A. Fischer \at Department of Astronomy, Yale University, New Haven, CT 06511 USA \email{debra.fischer@yale.edu}}
%
%
\titlerunning{55Cnc}
\maketitle

\abstract{The star 55 Cancri was one of the first known exoplanet hosts and each of the planets 
in this system is remarkable. Planets b and c are in a near 1:3 resonance. Planet d has a 14.5 year 
orbit, and is one of the longest known orbital periods for a gas giant planet. Planet e has a mass of 8 $M_\oplus$ 
and transits this bright star, providing a unique case for modeling of the interior structure and atmospheric 
composition of an exoplanet. Planet f resides in the habitable zone of the star.  If the planets are approximately
co-planar, then by virtue of having one transiting planet, this is a system where the Doppler technique 
has essentially measured the true mass of the planets, rather than just $M \sin i$. The unfolding history 
of planet discovery for this system provides a good example of the challenges and importance of 
understanding the star to understand the planets. }

\section{Introduction}
Transiting exoplanets with well-determined masses offer a unique opportunity for understanding the interior structure, 
chemical composition, evolution and atmospheric properties of exoplanets. The brightest known transit host 
star is 55 Cancri (55 Cnc), with 
a $8 M_\oplus$ mass transiting planet in a 0.7356 day orbit \citep{Winn2011, Demory2011}. This star also has 4 other known
exoplanets. The gas giant planet 55 Cnc b \citep{Butler1997} was the fourth exoplanet discovered with the Doppler technique. 
Five years later, two more planets, 55 Cnc c and d, were announced in the system \citep{Marcy2002}.  55 Cnc d was remarkable 
in being the first planet known to orbit beyond 4AU. \citet{McArthur2004} announced the discovery of, 55 Cnc e. 
It was later learned that this orbital period was an alias of the true period, which is a remarkable 0.7365 days (18 hours).
This revision distinguished 55 Cnc~e as having the shortest known orbital period at that time. 
\citet{Fischer2008} added the last known planet to this system, 55 Cnc f, which orbits in the habitable zone. 

Over the decade of exoplanet discoveries in the 55 Cnc system, several things went wrong that led to transient periods of 
confusion and misinterpretations.  An incorrect infrared excess was measured, leading to the false interpretation 
of a significant debris disk. The modeling of an orbital period for 55 Cnc~e was a case study of aliases in time series data; that period was 
revised from 2.8~d \citep{McArthur2004} to 0.74~d \citep{DawsonFabrycky2010}. 
Before the revision to this orbital period for 55 Cnc~e, photometric surveys that targeted the incorrect transit ephemeris 
naturally missed what eventually turned out to be one of the most scientifically valuable transiting planets. 
The carbon abundance in the 
host star was intially overestimated, leading to confusion in the interpretation of the interior structure of 55 Cnc e. 
Ultimately, the story of 55 Cnc demonstrates the importance of understanding the host star to understand the orbiting planets.

\section{The Host Star}
Understanding the host star is important for understanding the nature (mass, radius, orbital evolution) of prospective 
planets. 55 Cancri (also known as 55 Cnc, $\rho^1$ Cancri, HD~75732, HIP~43587, HR~3522) is a bright G8V star, just 
41 parsecs away. It is the primary component of a visual binary system. The secondary component, $\rho^1 {\rm Cnc \,B}$, is 
about 7 magnitudes fainter ($V \sim 13$) with an angular separation of $\sim 85"$, corresponding 
to a projected physical separation of 1100 AU.  The absolute radial velocities for the primary and secondary star 
are $27.3 \pm 0.3$ and $27.4 \pm 0.3$ \kms\ respectively \citep{Nidever2002}, 
providing strong support for a gravitationally bound wide binary system. 

\subsection{The Confusing Question of a Debris Disk}
One early controversy over 55 Cnc centered on whether or not the star had a reprocessing disk.  
\citet{Dominik1998} suggested that 55 Cnc had a Vega-like dust disk based on Infrared Space Observatory (ISO) data and 
\citet{TrillingBrown1998} claimed that they had resolved the disk out to $3.''2$ with near infrared coronographic images.
However, \citet{Jayawardhana2000} was not able to confirm this result; they found the sub-mm emission to be a factor of 
100 lower than expected for the reported disk. \citet{Schneider2001} imposed an upper limit on the NIR flux with 
NICMOS on HST that was 10 times lower than that reported by \citet{TrillingBrown1998}. 
\citet{Jayawardhana2002} found 3 faint background sub-mm sources and it is possible that the faint binary companion 
may have contributed to the early source confusion. The best current assessment is that a reprocessing disk has not yet
been detected around 55 Cnc. 

\subsection{Piecing Together the Fundamental Parameters}
\citet{Ford1999} list a sampling of published effective temperatures 
for 55 Cnc that range from 4460 to 5336 K and discuss their inability to find a self-consistent stellar evolutionary model, given the poor 
quality of the observed parameters. Even the question of whether 55 Cnc was a main sequence star or a subgiant was disputed.
While 55 Cnc is a few tenths of a magnitude brighter than the zero age main sequence, this is likely because of high 
metallicity, which was first noted by \citet{GreensteinOinas1968}. 

Although stellar ages are notoriously difficult to determine \citep{MamajekHillenbrand2008}, model isochrones
can be used to estimate an age for 55 Cnc. Published photometry was fitted with a spectral energy distribution by \citet{vonbraun2011} 
to derive a bolometric luminosity. Using their interferometrically measured stellar radius $0.943 \pm 0.01$ to   
derive \teff\ = $5196 \pm 24 {\rm K}$, the authors adopted the ${\rm [Fe/H]}$ published by \citet{VF05} to fit Yonsei-Yale 
isochrones \citep{Demarque2004, Kim2002, Yi2004}, yielding a stellar mass of $0.905 \pm 0.015 M_\odot$ and an age 
of $10.2 \pm 2.2$ Gyr. This age was initially at odds 
with the statistical kinematic age of 1 Gyr suggested by \citet{Eggen1995}; however the kinematic age was later 
revised by \citet{Reid2002} who find the space motion velocity relative to the LSR to be 29.5 \kms, similar to disk stars 
with ages between $2 - 8$ Gyr. \citet{Gonzalez1998}, and \citet{Brewer2016} both carried out fine spectroscopic modeling to 
determine stellar temperature and abundances and then used isochrone fitting to estimate the stellar age of 7.9 Gyr for 55 Cnc.

The age, rotation rate, magnetic field strength and chromospheric activity of stars are all physically correlated and 
therefore can be used to collectively piece together a self-consistent characterization of a star.
As magnetic field lines emerge from the surfaces of stars, they suppress local convection and produce cooler 
regions, or starspots, in the photosphere. Photometric rotation periods can be measured 
for younger stars with larger spots (from stronger magnetic fields) using ground-based time series photometry. 
\citet{Marcy2002} obtained ground-based photometry for 55 Cnc that was constant at the level of 
0.0012 magnitudes without any indication of rotational variability, consistent with a slowly rotating star 
with small spots.

Chromospheric activity measurements offer the best way for estimating the rotational 
periods of slowly rotating stars like 55 Cnc. Magnetic fields transport energy into the lower chromosphere 
and produce emission in deep spectral line cores such as the calcium II H (396.9 nm) and K (393.4 nm), calcium infrared 
triplet (849.8, 845.2, 866.2 nm), and the hydrogen Balmer lines (486.1 and 656.3 nm). 
The correlation between chromospheric activity and the rotation period of stars was first noted by \citet{Kraft1967}.
The 30-year Mount Wilson Observatory HK Project \citep{Wilson1978, Vaughan1978, Duncan1991, Baliunas1995} 
measured Ca II H\&K emission to study stellar chromospheric activity and variability in bright stars. The consistency 
and duration of this data set make it a touchstone for all studies of chromospheric activity. 

Ca II H\&K emission is parameterized by $S_{HK}$, a measure of flux in the line core relative to 
flux in adjacent continuum bands. Since late type stars have much weaker near-UV continuum 
than earlier type stars, $S_{HK}$ values cannot be directly compared for stars 
with different spectral types. Instead, $S_{HK}$ can be transformed to $\log R'_{HK}$, the fraction 
of a star's bolometric luminosity from the lower chromosphere 
after accounting for photospheric contributions \citep{Noyes1984b, Soderblom1985, MamajekHillenbrand2008}.
The \citet{Noyes1984b} calibration between Ca II H\&K emission and rotation periods for stars with a range 
of $B - V$ color from 0.4 to 1.4 provides a good way to estimate rotation periods, even for older and more slowly rotating stars.
\citet{Henry2000} report an average $S_{HK}  = 0.19$ and $\log R'_{HK} = -4.949$ over 6 yr of monitoring 55 Cnc, 
implying a rotation period, ${\rm P_{rot}} = 42.2$d, consistent with other activity-based estimates of 42 to 44 
days \citep{Soderblom1985, Baliunas1997, BellBranch1976}. \citet{IsaacsonFischer2010} measure a 
slightly lower activity level, $\log R'_{HK}$ = -4.991, implying a rotational period of about 47 days. 

\begin{table}
\label{tab:55Cnc}       
\begin{tabular}{p{2.4cm}p{3cm}p{5.9cm}}
\hline\noalign{\smallskip}
Parameter & Value & Source  \\
\noalign{\smallskip}\svhline\noalign{\smallskip}
Spectral Type       &  G8V                & \citet[{\it Hipparcos}]{Perryman97}     \\
V Magnitude         &  6.1                  &  \citet[{\it Hipparcos}]{Perryman97}    \\
Parallax       & $81.03 \pm 0.75 \,{\rm mas}$ & \citet[{\it Hipparcos}]{Perryman97}     \\
${\rm M_V}$         & 5.64                &  (using {\it Hipparcos} photometry and parallax)     \\
Radial Velocity     & $27.3 \pm 0.3$ & \citet{Nidever2002}   \\
Mass                    & 0.71                 &  \citet{Brewer2016}   \\
Radius ($^a$iso)      & 0.93                 &  \citet{Ford1999}     \\
Radius  ($^a$iso)     & 0.92                 &  \citet{Brewer2016}     \\
Radius ($^a$intrfmtry)   & $0.943 \pm 0.01$    &  \citet{vonbraun2011}     \\
Luminosity     & 0.59 $L_\odot$      & \citet{Brewer2016}   \\
${\rm B - V}$       & 0.869               & \citet[{\it Hipparcos}]{Perryman97}  \\
${\rm T_{eff}}$ ($^a$intrfmtry)   & $0.943 \pm 0.01$    &  \citet{vonbraun2011}     \\
${\rm T_{eff}}$ ($^a$spectr)   & 5250 K    & \citet{Brewer2016}      \\
$\log g$ ($^a$iso)   & 4.5                 &  \citet{Ford1999}    \\
$\log g$ ($^a$spectr)   & 4.36    &  \citet{Brewer2016}    \\
${\rm [M/H]}$       & 0.34                 & \citet{Brewer2016}  \\
${\rm [Fe/H]}$      & 0.35                 & \citet{Brewer2016} \\
${\rm [Mg/H]}$    & 0.32                 & \citet{Brewer2016} \\
${\rm [Si/H]}$     & 0.33                 & \citet{Brewer2016} \\
${\rm [O/H]}$      & 0.32                  & \citet{Brewer2016} \\
Mg/Si                  & 0.87                & \citet{DelgadoMena2010} \\
Mg/Si                  & 0.984                & \citet{BF2016} \\
Si/Fe                   & 1.098                & \citet{Brewer2016}  \\
${\rm C/O}$       & 1.12                & \citet{DelgadoMena2010}  \\
${\rm C/O}$        & $0.534 \pm 0.06$  & \citet{BF2016} \\
${\rm P_{rot}}$ ($S_{HK}$)  &  42.2d & \citet{Henry2000}   \\
${\rm P_{rot}}$ ($S_{HK}$)  &  42.2d & \citet{IsaacsonFischer2010}   \\
$S_{HK}$            & 0.19                   &  \citet{Henry2000}  \\
$S_{HK}$            & 0.18                   &  \citet{IsaacsonFischer2010}  \\
$\log R^\prime_{HK}$  & -4.949       & \citet{Henry2000}  \\
$\log R^\prime_{HK}$  & -4.991       & \citet{IsaacsonFischer2010}  \\
Age  ($^a$iso)  & $10 \pm 2.5$ Gyr     & \citet{vonbraun2011} \\
Age  ($^a$iso)   & $8.4^{+7.1}_{-8.3}$ Gyr  & \citet{Ford1999} \\
Age  ($^a$iso)   & 7.9 Gyr               & \citet{Gonzalez1998} \\
Age  ($^a$iso)    & 7.9 Gyr               & \citet{Brewer2016} \\
Age  ($^a$kin)   & 2 - 8 Gyr             & \citet{Eggen1995} \\
Age  ($^a$kin)   & 2 - 8 Gyr             & \citet{Reid2002} \\
Age  ($^a$act)   & $6.44 \pm 1$ Gyr   & \citet{IsaacsonFischer2010} \\
\noalign{\smallskip}\hline\noalign{\smallskip}
\end{tabular}
$^a$Abbreviations: iso: isochrone models; intrfmtry: interferometric measurement; spectr: spectroscopic analysis; kin: kinematic evidence; 
act: inferred by chromospheric activity 
\end{table}

The Mount Wilson HK survey also demonstrated a relation between stellar rotation periods and the magnetic cycle periods, 
analogous to the 11-year solar magnetic cycle. Using the relation derived by \citet{Noyes1984a} and \citet{Noyes1984b}: 
$P_{cyc} \sim \left(P_{rot} / \tau_c \right)^{1.25}$, where $P_{cyc}$ 
is magnetic cycle period, $P_{rot}$ is the rotation period, and $\tau$ is the  turnover time at the base of the convective 
zone, the estimated period for the magnetic cycle of 55 Cnc is $9-10$ years. 

Young stars rotate faster and have stronger magnetic fields, empirically suggesting a causal 
relation between stellar rotation and the magnetic dynamo. With time, the rotation period
of a star slows down. The spin-down relation can be inverted to estimate the age of a star, 
given its rotation period. \citet{Donahue1998} and \citet{Henry2000} used the chromospheric activity 
measurement from Ca II H\&K emission to estimate an activity-based stellar age of $6.44 \pm 1$ Gyr,
a good match to ages derived by evolutionary models discussed above. 

The fundamental data for 55 Cnc has improved significantly over time. The interferometric stellar radius measurement 
\citep{vonbraun2011} is secure and the spectroscopic analysis is now both more accurate \citep{Brewer2015} and 
more precise, yielding an effective temperature of $5250 \pm 35 {\rm K}$, $\log g = 4.36 \pm 0.05$, 
and ${\rm [Fe/H] = 0.35}$ \citep{Brewer2016}. The best current fundamental stellar parameters for 55 Cnc are 
compiled in Table~\ref{tab:55Cnc}. 

\section{Unfolding Exoplanet Discoveries}
The first detected exoplanet orbiting 55 Cnc was a gas giant in a 14.65 day orbit \citep{Butler1997}, found with 
the Doppler technique. Alternative explanations for the radial velocity signal, including star spots and pulsations, 
were considered by the authors. However, spots would exhibit the same $\sim 44$ day rotation period as the star 
and pulsations would have to produce a photometrically detectable 7.3\% change in radius to explain 
the 77 \ms\ velocity amplitude.   

Five years later, \citet{Marcy2002} announced two more planets: 55 Cnc~c, with an orbital period of 44.276 d and 
mass $M \sin i = 0.21 M_{Jup}$, and a gas giant planet, 55 Cnc~d with an orbital period of $5360 \pm 400$ days.  
The best 3-planet Keplerian model still had a reduced chi-squared fit that seemed too high with a residual radial 
velocity rms of 8.5 \ms.  Because the period of 55 Cnc~c was so close to the rotational period of the star, the 
authors discussed the possibility of star spots as the orgin for the signal.  However, two compelling points were 
made: star spots fade after a month or two, but the signal had persisted with unfailing regularity for almost 14 years. 
In addition, the star was chromospherically inactive and photometrically stable. Noting the closeness of planets b 
and c to a 3:1 resonance, exhaustive dynamical simulations were carried out; no measurable non-Keplerian 
perturbations were found that might account for the higher than expected residual scatter. 

\citet{McArthur2004} combined data from the Hobby Eberly Telescope High-Resolution Spectrograph with 
the published Lick Hamilton data and announced the detection of 55 Cnc e, a Neptune-mass planet with 
an orbital period of 2.808d. The small velocity amplitude, $K=6.67 \pm 0.81$ \ms, corresponded to a planet 
with a mass similar to Neptune.  \citet{McArthur2004} highlighted the \citet{Bodenheimer2001} suggestion 
that low mass close-in planets could experience significant heating from tidal interactions that 
inflate the planet radius and result in loss of a significant fraction of the atmospheric mass, leaving behind 
a rocky core. While the orbital period for 55 Cnc~e would later turn out to be an alias of the true signal, inclusion 
of a fourth Keplerian signal helped to reduce scatter in the residual velocities.  \citet{McArthur2004} also analyzed 
data from the HST FGS and reported a tentative detection of an astrometric signal for 55 Cnc~d that implied 
an inclination of $37^\circ \pm 7^\circ$. If correct, this would imply a substantial non-coplanarity 
relative to 55 Cnc~e, which was later observed to transit. 

\subsection{The Mysterious Case of Aliases}

Shortly after the McArthur publication appeared, \citet{Wisdom2005} raised doubts about the 2.8-d planet, 55 Cnc e, 
and argued that the signal might be an alias of the 43.93-day planet. Wisdom found an additional 
signal, a sinusoid with amplitude 3.15 m/s and period of 261 days. This important work was not published in a 
peer-reviewed journal; however, a brief manuscript is posted on Jack Wisdom's website.  

A few years later, \citet{Fischer2008} provided an 18-year update on the 55 Cnc system. The authors confirmed 
the four previous planets (including the spurious period for 55 Cnc e) as well as the fifth planet, 55 Cnc f with 
an orbital period of 260 days and mass, $M\ sin i = 45.7 M_\oplus$. This paper improved the orbital parameters 
for the longest period orbit. The authors note that even after fitting for these five planets, the residual radial 
velocity scatter still seemed excessive. They speculated that this might be caused by systematic errors (from the 
instrument or the stellar photosphere), underestimated formal RV errors, or additional low amplitude planets.  
Their simulations showed that a dynamically stable sixth planet would have eluded detection in the following parameter 
space of mass and period: $Msini < 50 M_\oplus$ with orbital periods between 300 and 850 days, $Msini < 100 M_\oplus$ for periods from 850 to 1500 days, and $Msini < 250 M_\oplus$ for periods from 1750 to 4000 days. These regions of parameter space remain interesting 
hunting grounds for additional planets in the 55 Cnc system. 

In 2010, Dawson \& Fabrycky outlined a cookbook approach for distinguishing aliases in undersampled time 
series data. Indeed, as \citet{DawsonFabrycky2010} suspected, 1-day aliases were common enough in time 
series radial velocity data sets that a minimum period search for periodogram analysis was often set to be just 
longer than one day. \citet{DawsonFabrycky2010} investigated the published data 
sets and found that the model for some Doppler-detected planets, including 55 Cnc e, was affected by 
daily aliases. They revised the orbital period first found by \citet{McArthur2004} from 2.8 days to 0.7365 days. 
At the time, this was the shortest orbital period of any known exoplanet.

\begin{table}
\caption{Detected Planets orbiting 55 Cnc}
\label{tab:55Cnc_planets}       
\begin{tabular}{p{0.8cm}p{2.5cm}p{1.6cm}p{1.9cm}p{0.9cm}p{0.9cm}p{0.9cm}p{0.9cm}}
\hline\noalign{\smallskip}
            & Period                             &    ${\rm T_p}$   &                         & $\omega$   &  K       & Mass                    &  ${\rm a_{rel}}$   \\
Planet & [days]                               &      [JD]             &   Eccentricity    &   [deg]  &  \ms   &  ${\rm M_\oplus}$   &  [AU]                  \\
\noalign{\smallskip}\svhline\noalign{\smallskip}
b      &  14.65194 $\pm 0.00002$   &  2456830.92    &  0.013 $\pm 0.01$  & 138   &  70.9     &  258.4         &  0.114       \\
c      &  44.392     $\pm 0.0012$     &  2453973.66    &  0.082 $\pm 0.02$  &  18.5  &  10.6    &  55.6           & 0.24          \\
d      &  5285        $\pm 4.5$           &  2454390.96    &  0.032 $\pm 0.1  $  &  357   &  45.2    & 1170           & 5.8           \\
e      &  0.73655   $\pm 5.1e-7$      &  2459676.49    &  0.22   $\pm 0.05$  &  156   &  5.7      & 7.5              & 0.0156      \\
f      &  259.4        $\pm 0.1$           &  2459416.62    &  0.27   $\pm 0.05$  &  305   &  4.2      & 38.3            & 0.78           \\
\noalign{\smallskip}\hline\noalign{\smallskip}
\end{tabular}
\end{table}

The Keplerian signals for the five 55 Cnc planets have been refit here, 
using published radial velocities from Lick Observatory \citep{Fischer2014} and Keck Observatory \citep{Butler2017}. 
For each individual planet, the theoretical Keplerian velocities are overplotted on the observed 
data after subtracting off Keplerian velocities derived for the other planets using the best fit parameters listed 
in Table~\ref{tab:55Cnc_planets}. \citet{Nelson2014} carried out a self-consistent Bayesian analysis of the 55 Cnc system 
with n-body integrations to investigate gravitational interactions between the planets. The Keplerian parameters modeled here 
agree within uncertainties with the \citet{Nelson2014} values except that we fit a slightly longer period for 55 Cnc~d  
and derive a smaller velocity amplitude and therefore a smaller mass for 55 Cnc~e. 

\begin{figure}
	\centering
	\begin{minipage}{0.48\textwidth}
		\centering
		\includegraphics[width=0.96\textwidth]{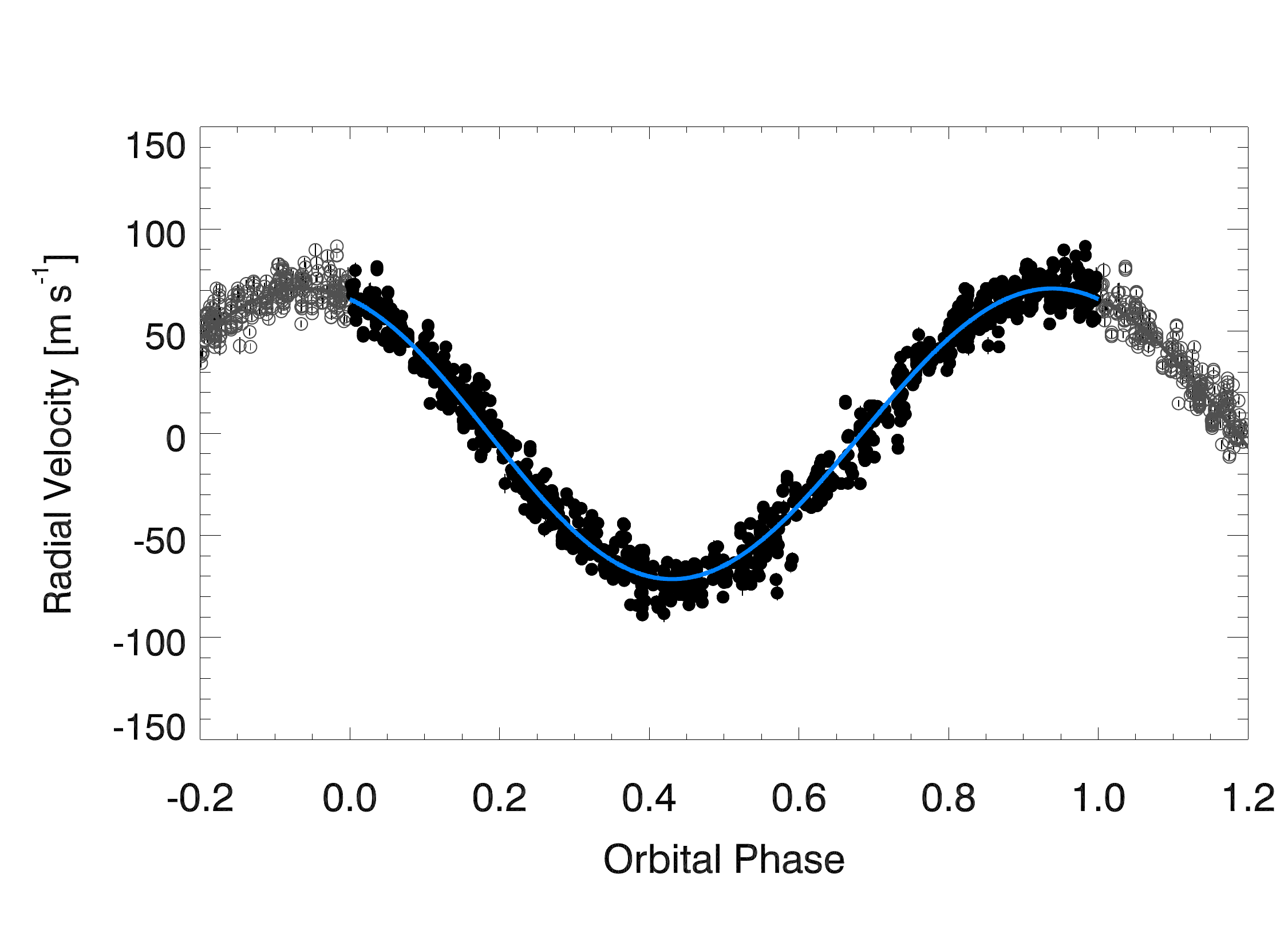}
		\caption{55 Cnc b, phase-folded at 14.65194 days. The amplitude of 70.9 \ms\ 
		implies a planet with $M \sin i = 0.81 M_{Jup}$. Theoretical curve is overplotted as a blue line. }
		\label{fig:fig1}      
	\end{minipage}\hfill
	\begin{minipage}{0.48\textwidth}
		\centering
		\includegraphics[width=0.96\textwidth]{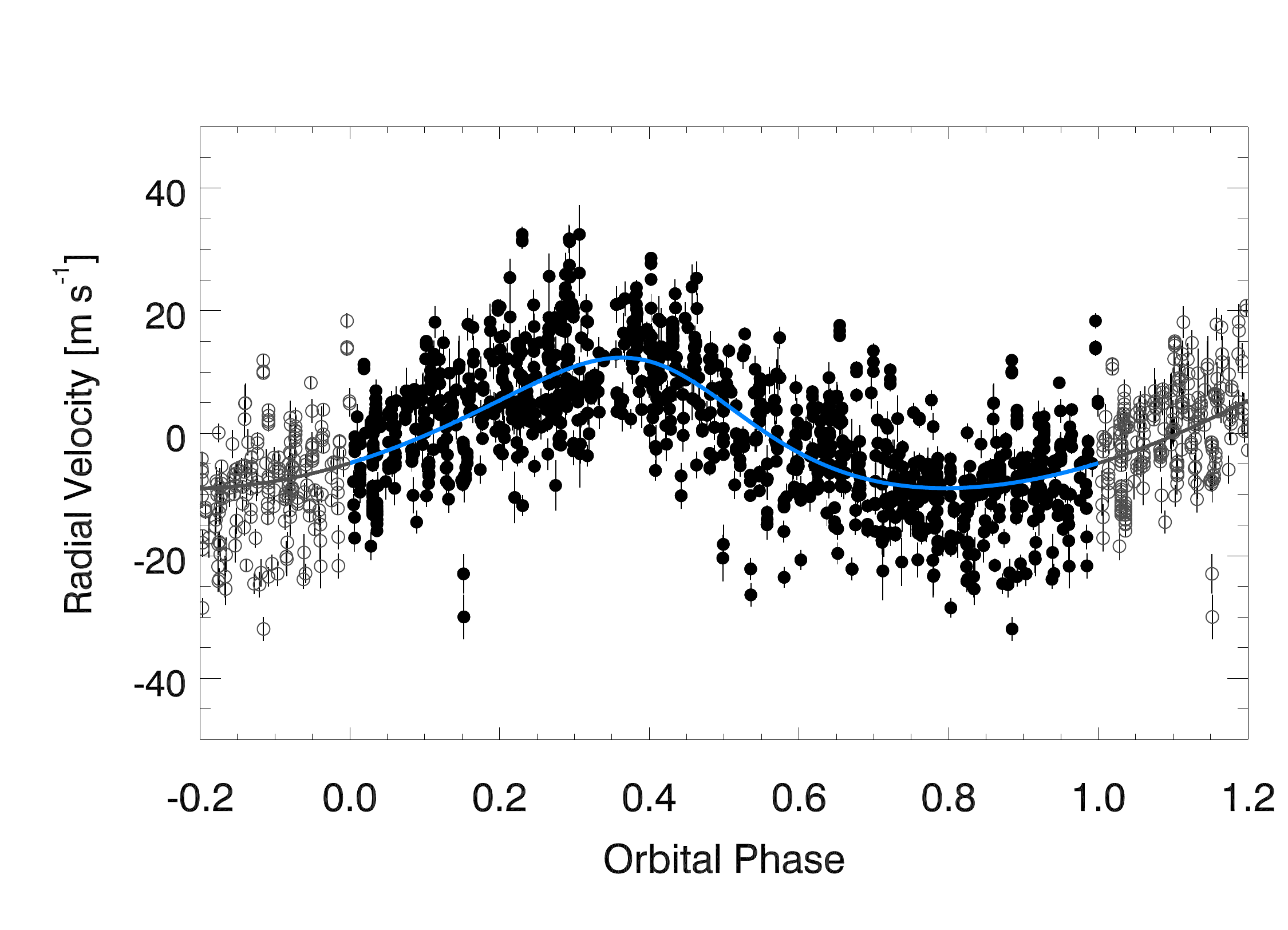}
		\caption{Radial velocities for 55 Cnc c, phase-folded at 44.392 days.
		The 10.6 \ms\ amplitude corresponds to a $ M \sin i = 55.6 M_\oplus$ planet. Model fit is overplotted as a blue line.}
		\label{fig:fig2}       
	\end{minipage}	
\end{figure}

\begin{figure}
	\centering
	\begin{minipage}{0.48\textwidth}
		\centering
		\includegraphics[width=0.96\textwidth]{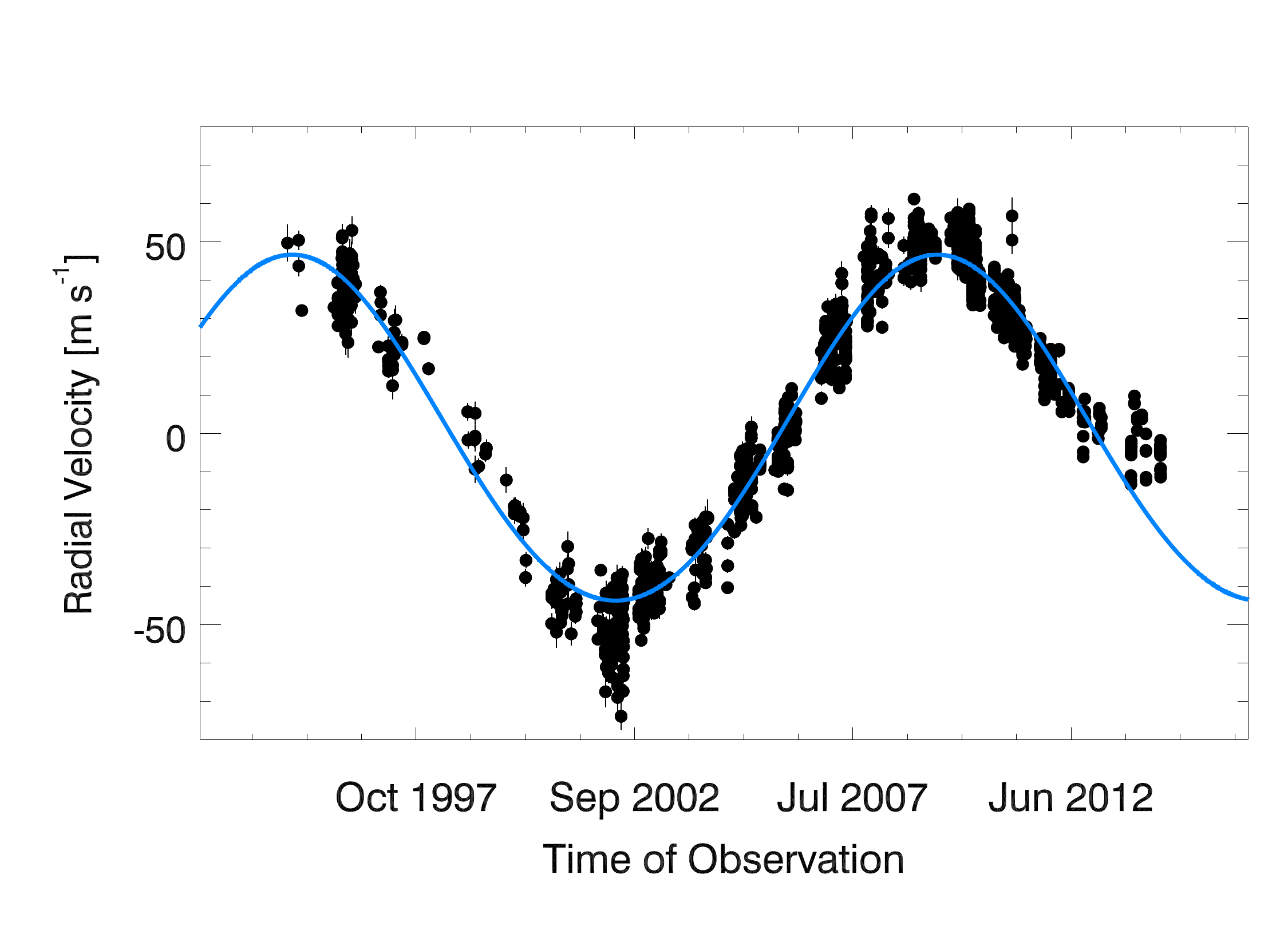}
		\caption{The time series data for 55 Cnc d with other planets removed, corresponds to a planet 
		with an orbital period of 14.5 years and $M \sin i = 3.7 M_{Jup}$. Keplerian fit is overplotted as a blue line.}
		\label{fig:fig3}       
	\end{minipage}\hfill
	\begin{minipage}{0.48\textwidth}
		\centering
		\includegraphics[width=0.96\textwidth]{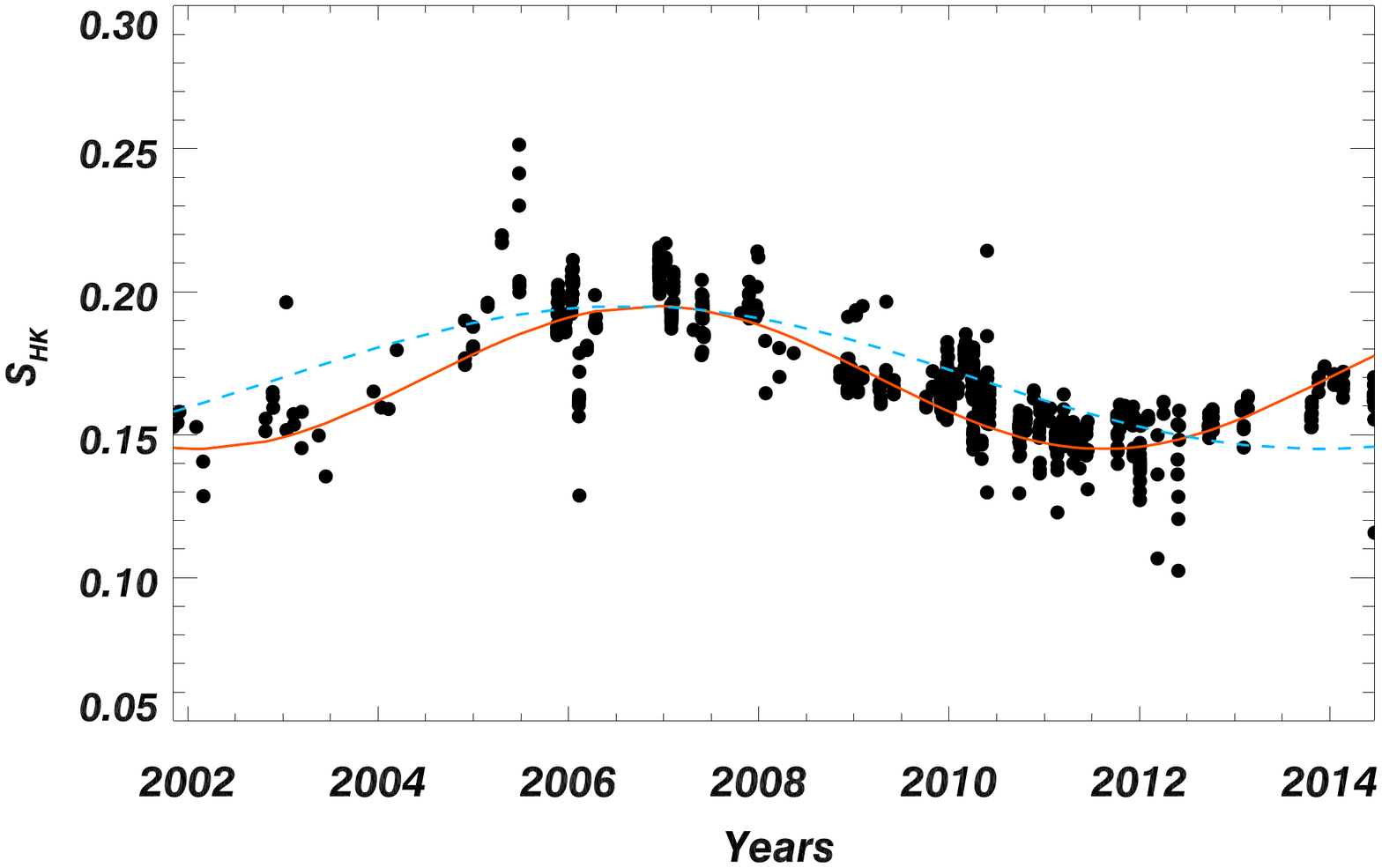}
		\caption{The magnetic activity cycle of 55 Cnc appears in the time series Ca II H\&K emission from the Keck spectra. 
		Consistent with the predicted correlation, the best fit period for $P_{cyc}$ is a 10 year period (red sinusoidal line). The 14.5 year 
		period of the outer planet (rescaled and shown as a blue dashed line) is not in phase with the activity cycle. }
		\label{fig:fig4}       
	\end{minipage}	
\end{figure}

\begin{figure}
	\centering
	\begin{minipage}{0.48\textwidth}
		\centering
		\includegraphics[width=0.96\textwidth]{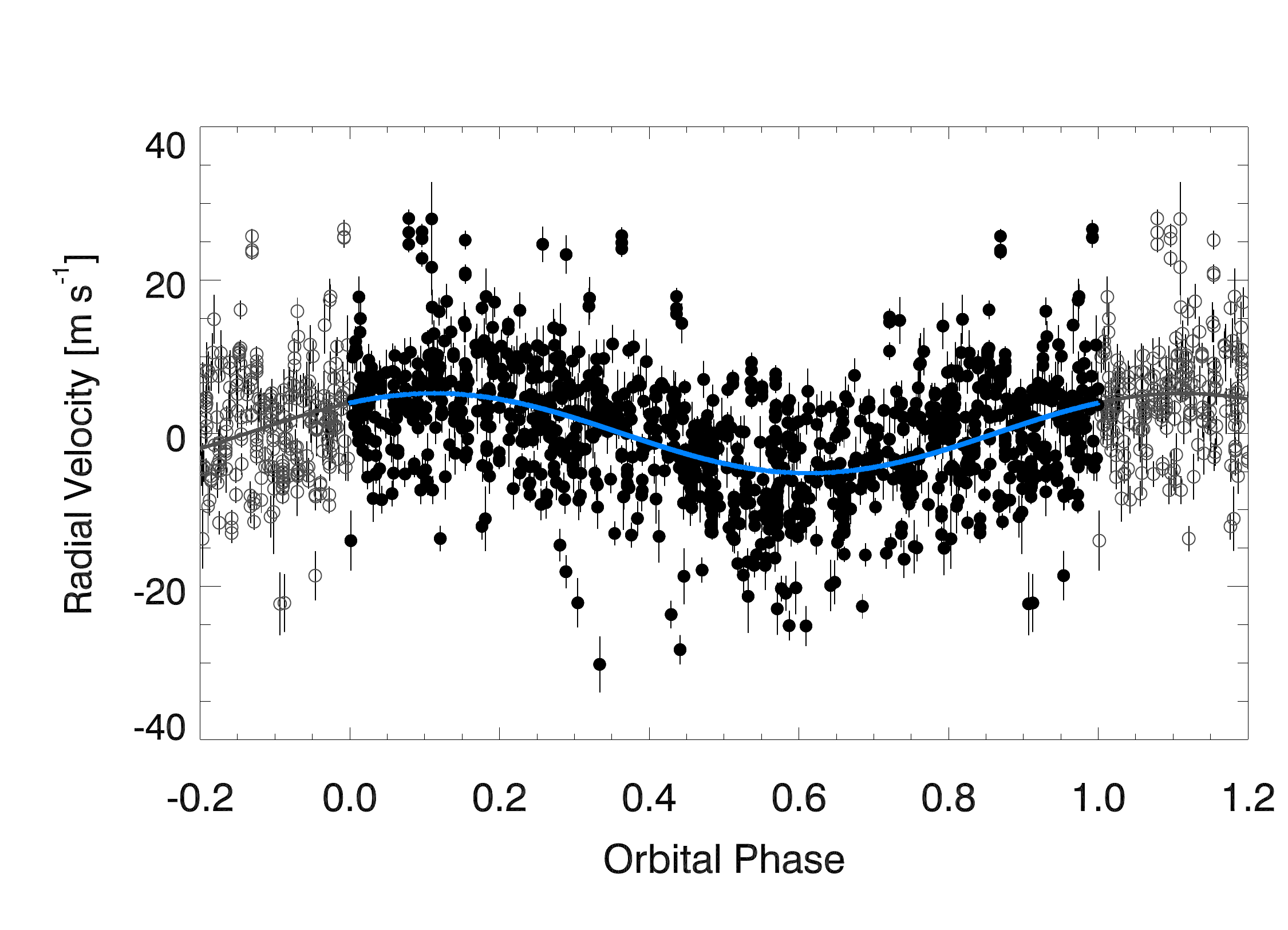}
		\caption{The 55 Cnc velocities  
		are phase-folded at 0.73655 days revealing 5.7 \ms\ signal that is modeled as 55 Cnc e, a $7.5 M_\oplus$ planet.
		Model fit is overplotted as a blue line.}	
		\label{fig:fig5}       
	\end{minipage}\hfill
	\begin{minipage}{0.48\textwidth}
		\centering
		\includegraphics[width=0.96\textwidth]{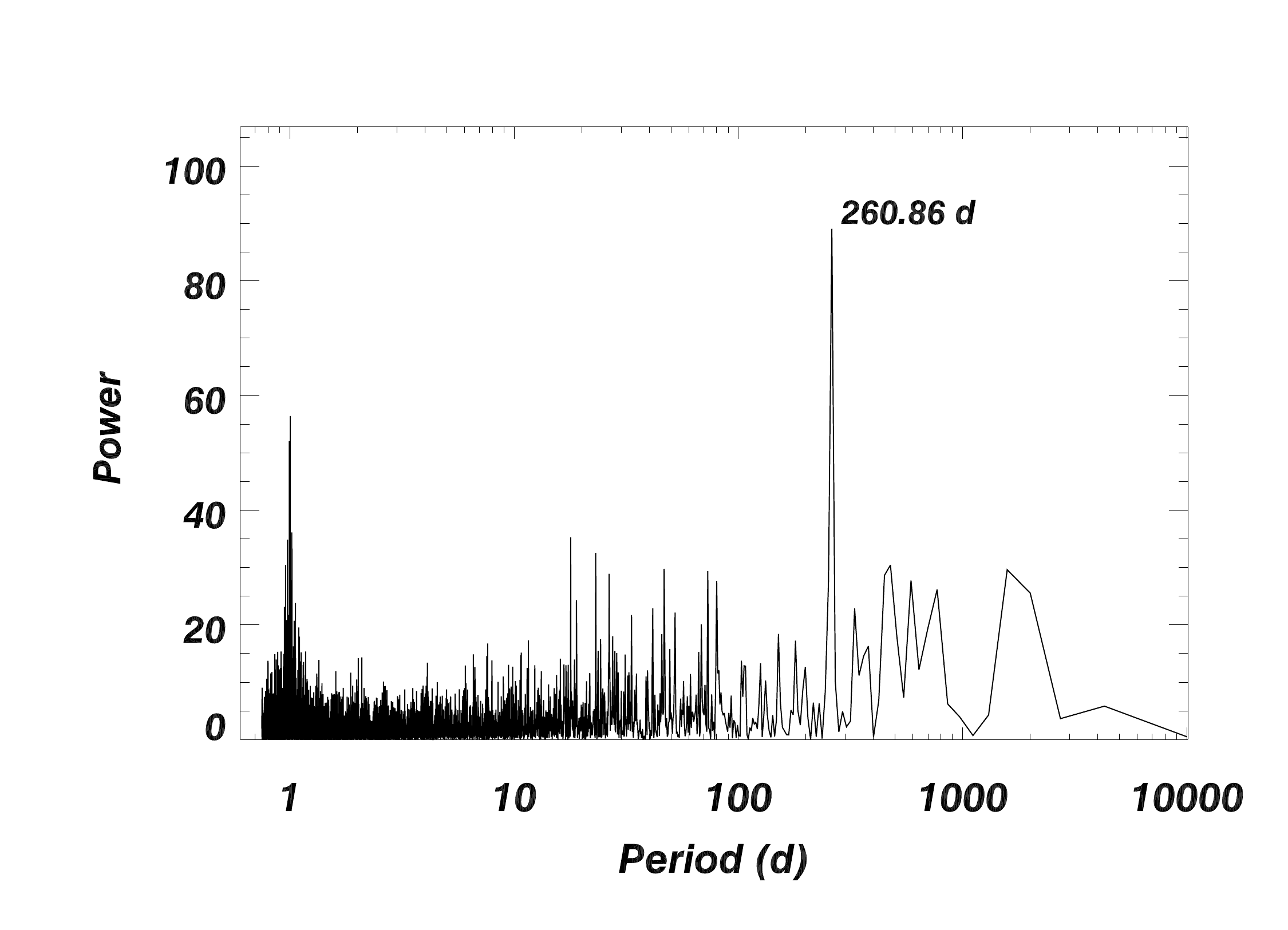}
		\caption{After fitting for planets b, c, d, e, a periodogram of the residual velocities shows a strong peak at 260 d.}
		\label{fig:fig6}       
	\end{minipage}	
\end{figure}

\begin{figure}
	\centering
	\begin{minipage}{0.48\textwidth}
		\centering
		\includegraphics[width=1.0\textwidth]{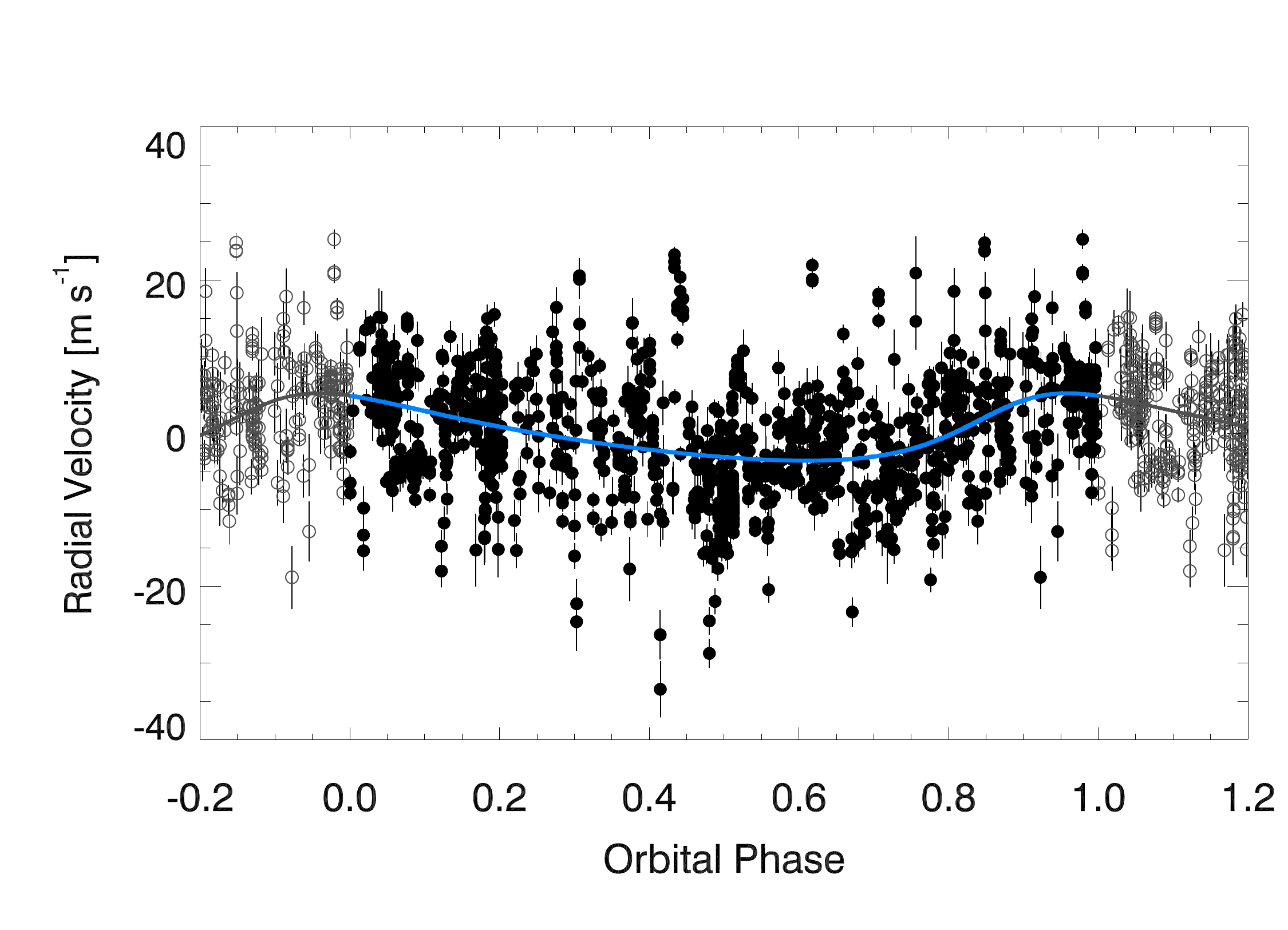}
		\caption{The radial velocity data are phase-folded 
		at the best fit period of 259.4 days for 55 Cnc f. Theoretical fit is overplotted as a blue line.}
		\label{fig:fig7}	
	\end{minipage}\hfill
	\begin{minipage}{0.48\textwidth}
		\centering
		\includegraphics[width=0.96\textwidth]{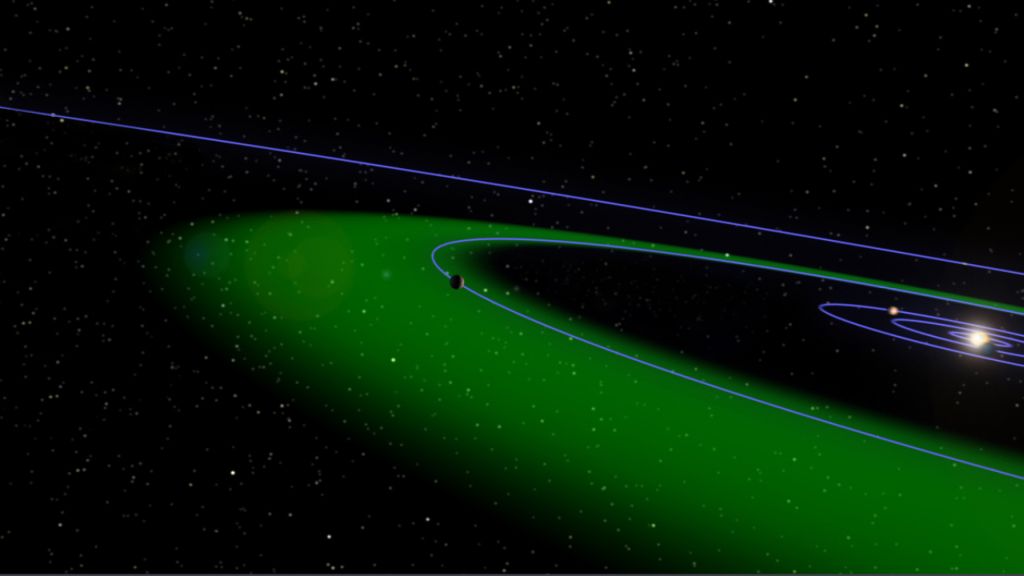}
		\caption{Artists rendition of the 55 Cnc exoplanet architecture, with 55 Cnc f in the habitable zone of the star.}
		\label{fig:fig8}       
	\end{minipage}	

\end{figure}

Figures \ref{fig:fig1} and \ref{fig:fig2}
show the best-fit, phase-folded velocities for 55 Cnc~b and c with the theoretical curve overplotted. 
Figure~\ref{fig:fig3} shows the RV data overplotted with a best-fit model with a period of 14.5 years for 55 Cnc~d. 
This long period is close enough to $P_{c}$, the magnetic cycle period, that it 
is worth considering whether this long period signal is the result of globally varying velocity fields induced during the 
magnetic cycle of 55 Cnc. \citet{Butler2017} also publish $S_{HK}$ values for the Keck data and this time series is shown 
in Figure~\ref{fig:fig4}. Overplotted on the $S_{HK}$ data is a scaled sinusoid (solid red line) with a 10-year period, consistent with 
the magnetic cycle period estimated above.  Also shown for reference is a scaled sinusoid (dashed blue line) with the same period as 
55 Cnc d. The inconsistency between the best fits for $P_{cyc}$ and the orbital period for 55 Cnc~d support the planetary interpretation 
for the 14.5 year signal. Figure~\ref{fig:fig5} showse the phase-folded data and theoretical fit for 55 Cnc~e. After fitting for planets b, c, d, and e,
a periodogram of the residuals (Figure~\ref{fig:fig6}) shows strong power at the period of 55 Cnc~f. The residual velocities are phase-folded
at this period and overplotted with the theoretical fit in Figure~\ref{fig:fig7}.  Figure 8 shows a artists rendition of the inner planetary system
with the orbit of 55 Cnc~f located in the habitable zone of the host star. 

\section{A Transiting Planet}
The change in orbital period from 2.8 days \citep{McArthur2004} to 0.74 days \citep{DawsonFabrycky2010} doubled the 
probability of detecting a transit. \citet{Demory2011} estimated a transit probability of 29\% for the revised orbital period and 
observed 55 Cnc at the predicted transit time on 6 January 2011 (Figure~\ref{fig:fig9}) using the IRAC 4.5 micron channel on 
the warm Spitzer space telescope. A month later, \citet{Winn2011} carried out nearly continuous photometric monitoring of the system 
with the {\it MOST} ({\it Microvariability and Oscillations of Stars}) Canadian satellite from 2011 February 7 to 22. Both teams 
detected a transit at the time predicted with the shorter orbital period and adopted the careful interferometric measurement 
of the stellar angular diameter using the CHARA array \citep{vonbraun2011} to derive a radius for the planet. 

Winn et al. 2011 modeled their phase-folded data (Figure~\ref{fig:fig10}) and measured a transit depth of $380 \pm 52$ ppm, implying a 
planet radius of $2.00 \pm 14 R_\oplus$. Adopting the stellar mass of $0.963^{+0.051}_{-0.029}$ from \citep{Takeda2007} 
and stellar radius of $0.943 \pm 0.01 \, R_\odot$ they derived a planetary mass of $8.63 M_\oplus$. \citet{Gillon2012} 
combined {\it MOST} observations with their new 4.5 $\mu$m {\it Warm Spitzer} IRAC data to derive a slightly larger 
radius of $2.17 \pm 0.10 R_\oplus$. This slightly larger radius leads them to conclude that 55 Cnc e has a gaseous envelope 
overlying a rocky nucleus. Given the proximity of the planet to the host star, the authors note that a plausible composition
for the envelope is water in super-critical form.  While the photometric errors from ground-based facilities are larger 
because of atmospheric scintillation, \citet{demooij2014} were able to detect the transit using ALFOSC with the 
2.5-m Nordic Optical Telescope.

\begin{figure}
	\centering
	\begin{minipage}{0.48\textwidth}
		\centering
		\includegraphics[width=0.96\textwidth]{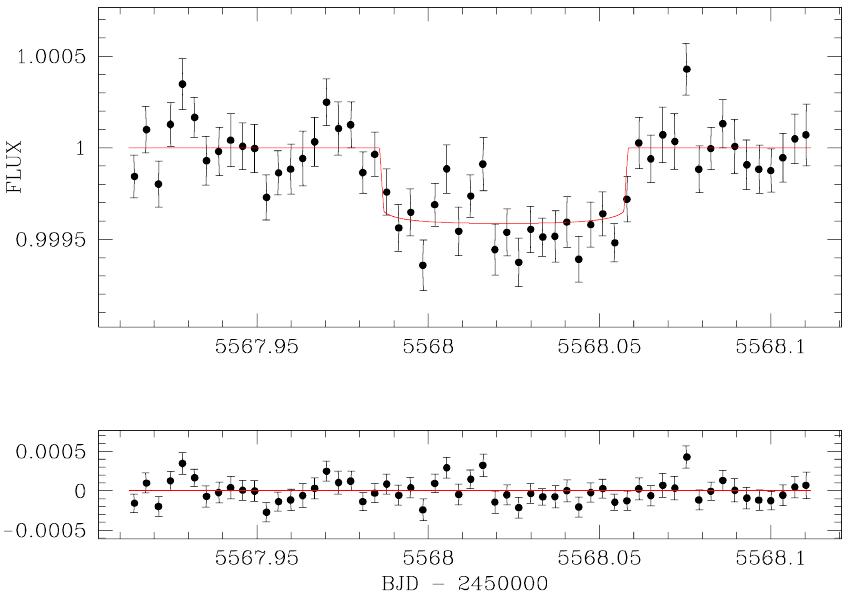}
		\caption{\citet{Demory2011} Spitzer transit of 55 Cnc e}
		\label{fig:55Cnce_Spitzertrans}       
	\end{minipage}\hfill
	\begin{minipage}{0.48\textwidth}
		\centering
		\includegraphics[width=0.96\textwidth]{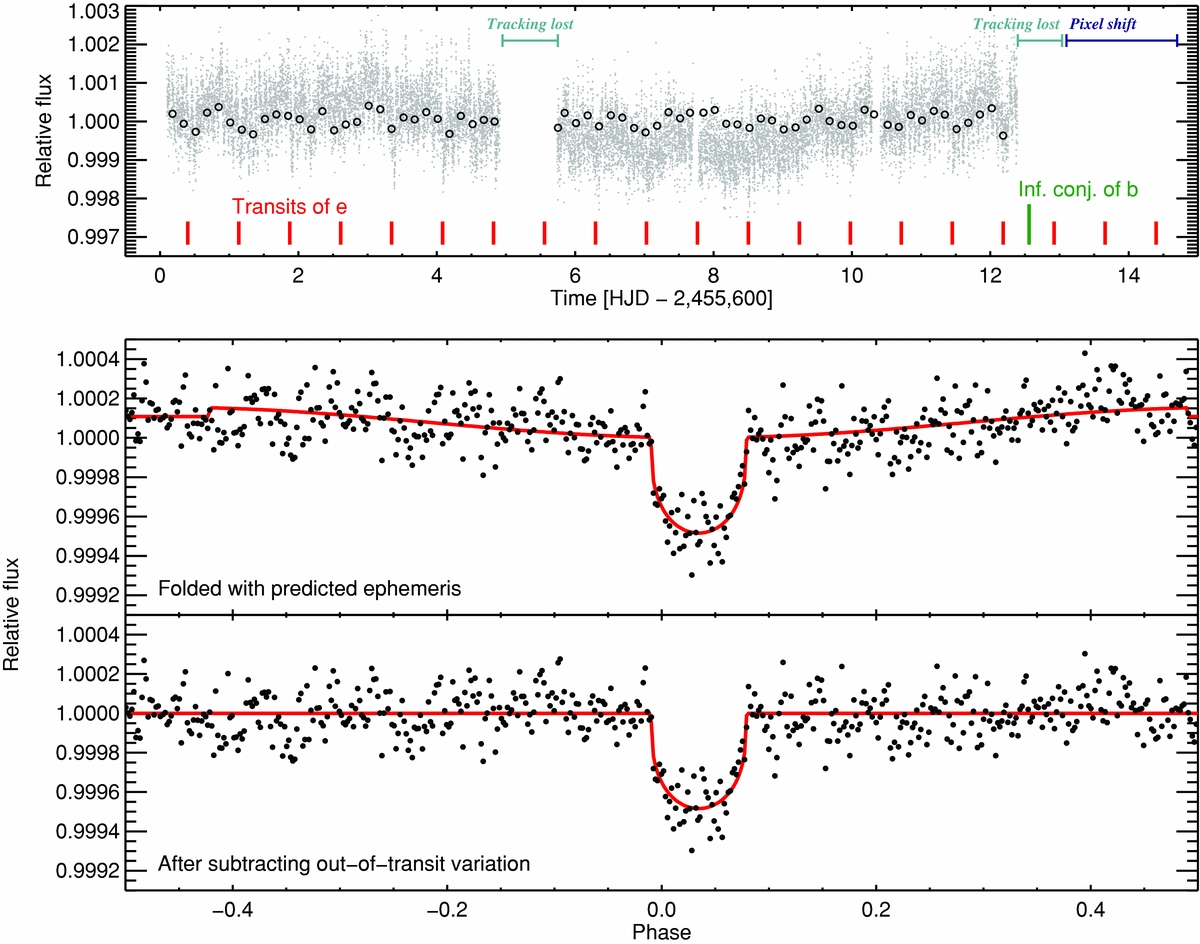}
		\caption{\citet{Winn2011} transit light curve from the MOST satellite, phase folded at P=0.74 d.}
		\label{fig:55Cnce_MOSTtrans}       
	\end{minipage}	
\end{figure}

\subsection{The Mistaken Identity of 55 Cnc~e}
With revised orbital parameters and a secure mass and radius for 55 Cnc~e, the characterization of the planet has been the subject 
of lively debate.  In particular, the ratio of carbon to oxygen in the protoplanetary disk influences the 
composition and structure of exoplanets. If ${\rm C/O}$ is above a threshold value, then carbonates 
will dominate \citep{BF2016, Madhu2012b}. Conversely, the planet composition will be primarily magnesium silicates 
if ${\rm C/O}$ is below that threshold. The nominal threshold value is $\sim 1$ at the formation site of 
planetesimals, but with planetesimal migration, the threshold value could be as low as 0.6 \citep{Moriarty2014}. 
The stellar abundance ${\rm C/O}$ ratio is a reasonable estimate of the volatile gas ratio in the protoplanetary
disk, and early estimates of ${\rm C/O}$ for 55 Cnc were as high as 1.12 \citep{DelgadoMena2010} leading 
\citet{Madhu2012a} to consider the possibility of a carbon-rich interior structure for 55 Cnc~e. 
The stellar ${\rm C/O}$ ratio has since been revised \citep{Brewer2016} and \citet{Demory2016b} provide 
a revised model for the interior structure of 55 Cnc~e that is consistent with a iron core and silicate mantle, 
similar to Earth.  

\subsection{The Future}
Our understanding of the planetary system orbiting 55 Cnc is likely to be revised again in the future. There is an 
enormous gap between 55 Cnc~f orbiting at 0.78 AU and 55 Cnc~d orbiting at 5.8 AU. If Nature abhors a vacuum, then 
there are sure to be additional planets awaiting discovery. 

\begin{acknowledgement}
Fischer acknowledges NASA grants NNX08AF42G and NNX15AF02G and thanks Yale University for 
research time that supported writing of this manuscript. 
\end{acknowledgement}

\bibliographystyle{aasjournal}  
\bibliography{HBexo_Fischer} 

\begin{thebibliography}{}
\expandafter\ifx\csname natexlab\endcsname\relax\def\natexlab#1{#1}\fi
\providecommand{\url}[1]{\href{#1}{#1}}
\providecommand{\dodoi}[1]{doi:~\href{http://doi.org/#1}{\nolinkurl{#1}}}
\providecommand{\doeprint}[1]{\href{http://ascl.net/#1}{\nolinkurl{http://ascl.net/#1}}}
\providecommand{\doarXiv}[1]{\href{https://arxiv.org/abs/#1}{\nolinkurl{https://arxiv.org/abs/#1}}}

\bibitem[{{Baliunas} {et~al.}(1997){Baliunas}, {Henry}, {Donahue}, {Fekel}, \&
  {Soon}}]{Baliunas1997}
{Baliunas}, S.~L., {Henry}, G.~W., {Donahue}, R.~A., {Fekel}, F.~C., \& {Soon},
  W.~H. 1997, \apjl, 474, L119, \dodoi{10.1086/310442}

\bibitem[{{Baliunas} {et~al.}(1995){Baliunas}, {Donahue}, {Soon}, {Horne},
  {Frazer}, {Woodard-Eklund}, {Bradford}, {Rao}, {Wilson}, {Zhang}, {Bennett},
  {Briggs}, {Carroll}, {Duncan}, {Figueroa}, {Lanning}, {Misch}, {Mueller},
  {Noyes}, {Poppe}, {Porter}, {Robinson}, {Russell}, {Shelton}, {Soyumer},
  {Vaughan}, \& {Whitney}}]{Baliunas1995}
{Baliunas}, S.~L., {Donahue}, R.~A., {Soon}, W.~H., {et~al.} 1995, \apj, 438,
  269, \dodoi{10.1086/175072}

\bibitem[{{Bell} \& {Branch}(1976)}]{BellBranch1976}
{Bell}, R.~A., \& {Branch}, D. 1976, \mnras, 175, 25,
  \dodoi{10.1093/mnras/175.1.25}

\bibitem[{{Bodenheimer} {et~al.}(2001){Bodenheimer}, {Lin}, \&
  {Mardling}}]{Bodenheimer2001}
{Bodenheimer}, P., {Lin}, D.~N.~C., \& {Mardling}, R.~A. 2001, \apj, 548, 466,
  \dodoi{10.1086/318667}

\bibitem[{{Brewer} \& {Fischer}(2016)}]{BF2016}
{Brewer}, J.~M., \& {Fischer}, D.~A. 2016, \apj, 831, 20,
  \dodoi{10.3847/0004-637X/831/1/20}

\bibitem[{{Brewer} {et~al.}(2015){Brewer}, {Fischer}, {Basu}, {Valenti}, \&
  {Piskunov}}]{Brewer2015}
{Brewer}, J.~M., {Fischer}, D.~A., {Basu}, S., {Valenti}, J.~A., \& {Piskunov},
  N. 2015, \apj, 805, 126, \dodoi{10.1088/0004-637X/805/2/126}

\bibitem[{{Brewer} {et~al.}(2016){Brewer}, {Fischer}, {Valenti}, \&
  {Piskunov}}]{Brewer2016}
{Brewer}, J.~M., {Fischer}, D.~A., {Valenti}, J.~A., \& {Piskunov}, N. 2016,
  \apjs, 225, 32, \dodoi{10.3847/0067-0049/225/2/32}

\bibitem[{{Butler} {et~al.}(1997){Butler}, {Marcy}, {Williams}, {Hauser}, \&
  {Shirts}}]{Butler1997}
{Butler}, R.~P., {Marcy}, G.~W., {Williams}, E., {Hauser}, H., \& {Shirts}, P.
  1997, \apjl, 474, L115, \dodoi{10.1086/310444}

\bibitem[{{Butler} {et~al.}(2017){Butler}, {Vogt}, {Laughlin}, {Burt},
  {Rivera}, {Tuomi}, {Teske}, {Arriagada}, {Diaz}, {Holden}, \&
  {Keiser}}]{Butler2017}
{Butler}, R.~P., {Vogt}, S.~S., {Laughlin}, G., {et~al.} 2017, ArXiv e-prints.
\newblock \doarXiv{1702.03571}

\bibitem[{{Dawson} \& {Fabrycky}(2010)}]{DawsonFabrycky2010}
{Dawson}, R.~I., \& {Fabrycky}, D.~C. 2010, \apj, 722, 937,
  \dodoi{10.1088/0004-637X/722/1/937}

\bibitem[{{de Mooij} {et~al.}(2014){de Mooij}, {L{\'o}pez-Morales},
  {Karjalainen}, {Hrudkova}, \& {Jayawardhana}}]{demooij2014}
{de Mooij}, E.~J.~W., {L{\'o}pez-Morales}, M., {Karjalainen}, R., {Hrudkova},
  M., \& {Jayawardhana}, R. 2014, \apjl, 797, L21,
  \dodoi{10.1088/2041-8205/797/2/L21}

\bibitem[{{Delgado Mena} {et~al.}(2010){Delgado Mena}, {Israelian},
  {Gonz{\'a}lez Hern{\'a}ndez}, {Bond}, {Santos}, {Udry}, \&
  {Mayor}}]{DelgadoMena2010}
{Delgado Mena}, E., {Israelian}, G., {Gonz{\'a}lez Hern{\'a}ndez}, J.~I.,
  {et~al.} 2010, \apj, 725, 2349, \dodoi{10.1088/0004-637X/725/2/2349}

\bibitem[{{Demarque} {et~al.}(2004){Demarque}, {Woo}, {Kim}, \&
  {Yi}}]{Demarque2004}
{Demarque}, P., {Woo}, J.-H., {Kim}, Y.-C., \& {Yi}, S.~K. 2004, \apjs, 155,
  667, \dodoi{10.1086/424966}

\bibitem[{{Demory} {et~al.}(2016){Demory}, {Gillon}, {Madhusudhan}, \&
  {Queloz}}]{Demory2016b}
{Demory}, B.-O., {Gillon}, M., {Madhusudhan}, N., \& {Queloz}, D. 2016, \mnras,
  455, 2018, \dodoi{10.1093/mnras/stv2239}

\bibitem[{{Demory} {et~al.}(2011){Demory}, {Gillon}, {Deming}, {Valencia},
  {Seager}, {Benneke}, {Lovis}, {Cubillos}, {Harrington}, {Stevenson}, {Mayor},
  {Pepe}, {Queloz}, {S{\'e}gransan}, \& {Udry}}]{Demory2011}
{Demory}, B.-O., {Gillon}, M., {Deming}, D., {et~al.} 2011, \aap, 533, A114,
  \dodoi{10.1051/0004-6361/201117178}

\bibitem[{{Dominik} {et~al.}(1998){Dominik}, {Laureijs}, {Jourdain de Muizon},
  \& {Habing}}]{Dominik1998}
{Dominik}, C., {Laureijs}, R.~J., {Jourdain de Muizon}, M., \& {Habing}, H.~J.
  1998, \aap, 329, L53

\bibitem[{{Donahue}(1998)}]{Donahue1998}
{Donahue}, R.~A. 1998, in Astronomical Society of the Pacific Conference
  Series, Vol. 154, Cool Stars, Stellar Systems, and the Sun, ed. R.~A.
  {Donahue} \& J.~A. {Bookbinder}, 1235

\bibitem[{{Duncan} {et~al.}(1991){Duncan}, {Vaughan}, {Wilson}, {Preston},
  {Frazer}, {Lanning}, {Misch}, {Mueller}, {Soyumer}, {Woodard}, {Baliunas},
  {Noyes}, {Hartmann}, {Porter}, {Zwaan}, {Middelkoop}, {Rutten}, \&
  {Mihalas}}]{Duncan1991}
{Duncan}, D.~K., {Vaughan}, A.~H., {Wilson}, O.~C., {et~al.} 1991, \apjs, 76,
  383, \dodoi{10.1086/191572}

\bibitem[{{Eggen}(1995)}]{Eggen1995}
{Eggen}, O.~J. 1995, \aj, 109, 1327, \dodoi{10.1086/117365}

\bibitem[{{Fischer} {et~al.}(2014){Fischer}, {Marcy}, \&
  {Spronck}}]{Fischer2014}
{Fischer}, D.~A., {Marcy}, G.~W., \& {Spronck}, J.~F.~P. 2014, \apjs, 210, 5,
  \dodoi{10.1088/0067-0049/210/1/5}

\bibitem[{{Fischer} {et~al.}(2008){Fischer}, {Marcy}, {Butler}, {Vogt},
  {Laughlin}, {Henry}, {Abouav}, {Peek}, {Wright}, {Johnson}, {McCarthy}, \&
  {Isaacson}}]{Fischer2008}
{Fischer}, D.~A., {Marcy}, G.~W., {Butler}, R.~P., {et~al.} 2008, \apj, 675,
  790, \dodoi{10.1086/525512}

\bibitem[{{Ford} {et~al.}(1999){Ford}, {Rasio}, \& {Sills}}]{Ford1999}
{Ford}, E.~B., {Rasio}, F.~A., \& {Sills}, A. 1999, \apj, 514, 411,
  \dodoi{10.1086/306935}

\bibitem[{{Gillon} {et~al.}(2012){Gillon}, {Demory}, {Benneke}, {Valencia},
  {Deming}, {Seager}, {Lovis}, {Mayor}, {Pepe}, {Queloz}, {S{\'e}gransan}, \&
  {Udry}}]{Gillon2012}
{Gillon}, M., {Demory}, B.-O., {Benneke}, B., {et~al.} 2012, \aap, 539, A28,
  \dodoi{10.1051/0004-6361/201118309}

\bibitem[{{Gonzalez} \& {Vanture}(1998)}]{Gonzalez1998}
{Gonzalez}, G., \& {Vanture}, A.~D. 1998, \aap, 339, L29

\bibitem[{{Greenstein} \& {Oinas}(1968)}]{GreensteinOinas1968}
{Greenstein}, J.~L., \& {Oinas}, V. 1968, \apjl, 153, L91,
  \dodoi{10.1086/180228}

\bibitem[{{Henry} {et~al.}(2000){Henry}, {Baliunas}, {Donahue}, {Fekel}, \&
  {Soon}}]{Henry2000}
{Henry}, G.~W., {Baliunas}, S.~L., {Donahue}, R.~A., {Fekel}, F.~C., \& {Soon},
  W. 2000, \apj, 531, 415, \dodoi{10.1086/308466}

\bibitem[{{Isaacson} \& {Fischer}(2010)}]{IsaacsonFischer2010}
{Isaacson}, H., \& {Fischer}, D. 2010, \apj, 725, 875,
  \dodoi{10.1088/0004-637X/725/1/875}

\bibitem[{{Jayawardhana} {et~al.}(2000){Jayawardhana}, {Holland}, {Greaves},
  {Dent}, {Marcy}, {Hartmann}, \& {Fazio}}]{Jayawardhana2000}
{Jayawardhana}, R., {Holland}, W.~S., {Greaves}, J.~S., {et~al.} 2000, \apj,
  536, 425, \dodoi{10.1086/308942}

\bibitem[{{Jayawardhana} {et~al.}(2002){Jayawardhana}, {Holland}, {Kalas},
  {Greaves}, {Dent}, {Wyatt}, \& {Marcy}}]{Jayawardhana2002}
{Jayawardhana}, R., {Holland}, W.~S., {Kalas}, P., {et~al.} 2002, \apjl, 570,
  L93, \dodoi{10.1086/341101}

\bibitem[{{Kim} {et~al.}(2002){Kim}, {Demarque}, {Yi}, \&
  {Alexander}}]{Kim2002}
{Kim}, Y.-C., {Demarque}, P., {Yi}, S.~K., \& {Alexander}, D.~R. 2002, \apjs,
  143, 499, \dodoi{10.1086/343041}

\bibitem[{{Kraft}(1967)}]{Kraft1967}
{Kraft}, R.~P. 1967, \apj, 150, 551, \dodoi{10.1086/149359}

\bibitem[{{Madhusudhan}(2012)}]{Madhu2012b}
{Madhusudhan}, N. 2012, \apj, 758, 36, \dodoi{10.1088/0004-637X/758/1/36}

\bibitem[{{Madhusudhan} {et~al.}(2012){Madhusudhan}, {Lee}, \&
  {Mousis}}]{Madhu2012a}
{Madhusudhan}, N., {Lee}, K.~K.~M., \& {Mousis}, O. 2012, \apjl, 759, L40,
  \dodoi{10.1088/2041-8205/759/2/L40}

\bibitem[{{Mamajek} \& {Hillenbrand}(2008)}]{MamajekHillenbrand2008}
{Mamajek}, E.~E., \& {Hillenbrand}, L.~A. 2008, \apj, 687, 1264,
  \dodoi{10.1086/591785}

\bibitem[{{Marcy} {et~al.}(2002){Marcy}, {Butler}, {Fischer}, {Laughlin},
  {Vogt}, {Henry}, \& {Pourbaix}}]{Marcy2002}
{Marcy}, G.~W., {Butler}, R.~P., {Fischer}, D.~A., {et~al.} 2002, \apj, 581,
  1375, \dodoi{10.1086/344298}

\bibitem[{{McArthur} {et~al.}(2004){McArthur}, {Endl}, {Cochran}, {Benedict},
  {Fischer}, {Marcy}, {Butler}, {Naef}, {Mayor}, {Queloz}, {Udry}, \&
  {Harrison}}]{McArthur2004}
{McArthur}, B.~E., {Endl}, M., {Cochran}, W.~D., {et~al.} 2004, \apjl, 614,
  L81, \dodoi{10.1086/425561}

\bibitem[{{Moriarty} {et~al.}(2014){Moriarty}, {Madhusudhan}, \&
  {Fischer}}]{Moriarty2014}
{Moriarty}, J., {Madhusudhan}, N., \& {Fischer}, D. 2014, \apj, 787, 81,
  \dodoi{10.1088/0004-637X/787/1/81}

\bibitem[{{Nelson} {et~al.}(2014){Nelson}, {Ford}, {Wright}, {Fischer}, {von
  Braun}, {Howard}, {Payne}, \& {Dindar}}]{Nelson2014}
{Nelson}, B.~E., {Ford}, E.~B., {Wright}, J.~T., {et~al.} 2014, \mnras, 441,
  442, \dodoi{10.1093/mnras/stu450}

\bibitem[{{Nidever} {et~al.}(2002){Nidever}, {Marcy}, {Butler}, {Fischer}, \&
  {Vogt}}]{Nidever2002}
{Nidever}, D.~L., {Marcy}, G.~W., {Butler}, R.~P., {Fischer}, D.~A., \& {Vogt},
  S.~S. 2002, \apjs, 141, 503, \dodoi{10.1086/340570}

\bibitem[{{Noyes} {et~al.}(1984{\natexlab{a}}){Noyes}, {Hartmann}, {Baliunas},
  {Duncan}, \& {Vaughan}}]{Noyes1984a}
{Noyes}, R.~W., {Hartmann}, L.~W., {Baliunas}, S.~L., {Duncan}, D.~K., \&
  {Vaughan}, A.~H. 1984{\natexlab{a}}, \apj, 279, 763, \dodoi{10.1086/161945}

\bibitem[{{Noyes} {et~al.}(1984{\natexlab{b}}){Noyes}, {Weiss}, \&
  {Vaughan}}]{Noyes1984b}
{Noyes}, R.~W., {Weiss}, N.~O., \& {Vaughan}, A.~H. 1984{\natexlab{b}}, \apj,
  287, 769, \dodoi{10.1086/162735}

\bibitem[{{Perryman} {et~al.}(1997){Perryman}, {Lindegren}, {Kovalevsky},
  {Hoeg}, {Bastian}, {Cr{\'e}z{\'e}}, {Donati}, {Grenon}, {Grewing}, {van Leeu\
  wen}, {van der Marel}, {Mignard}, {Murray}, {Le Poole}, {Schrijver}, {Turon},
  {Arenou}, {Froeschl{\'e}}, \& {Petersen}}]{Perryman97}
{Perryman}, M.~A.~C., {Lindegren}, L., {Kovalevsky}, J., {et~al.} 1997, \aap,
  323, L49

\bibitem[{{Reid}(2002)}]{Reid2002}
{Reid}, I.~N. 2002, \pasp, 114, 306, \dodoi{10.1086/339257}

\bibitem[{{Schneider} {et~al.}(2001){Schneider}, {Becklin}, {Smith},
  {Weinberger}, {Silverstone}, \& {Hines}}]{Schneider2001}
{Schneider}, G., {Becklin}, E.~E., {Smith}, B.~A., {et~al.} 2001, \aj, 121,
  525, \dodoi{10.1086/318050}

\bibitem[{{Soderblom}(1985)}]{Soderblom1985}
{Soderblom}, D.~R. 1985, \aj, 90, 2103, \dodoi{10.1086/113918}

\bibitem[{{Takeda} {et~al.}(2007){Takeda}, {Ford}, {Sills}, {Rasio}, {Fischer},
  \& {Valenti}}]{Takeda2007}
{Takeda}, G., {Ford}, E.~B., {Sills}, A., {et~al.} 2007, \apjs, 168, 297,
  \dodoi{10.1086/509763}

\bibitem[{{Trilling} \& {Brown}(1998)}]{TrillingBrown1998}
{Trilling}, D.~E., \& {Brown}, R.~H. 1998, \nat, 395, 775,
  \dodoi{10.1038/27389}

\bibitem[{{Valenti} \& {Fischer}(2005)}]{VF05}
{Valenti}, J.~A., \& {Fischer}, D.~A. 2005, \apjs, 159, 141,
  \dodoi{10.1086/430500}

\bibitem[{{Vaughan} {et~al.}(1978){Vaughan}, {Preston}, \&
  {Wilson}}]{Vaughan1978}
{Vaughan}, A.~H., {Preston}, G.~W., \& {Wilson}, O.~C. 1978, \pasp, 90, 267,
  \dodoi{10.1086/130324}

\bibitem[{{von Braun} {et~al.}(2011){von Braun}, {Boyajian}, {ten Brummelaar},
  {Kane}, {van Belle}, {Ciardi}, {Raymond}, {L{\'o}pez-Morales}, {McAlister},
  {Schaefer}, {Ridgway}, {Sturmann}, {Sturmann}, {White}, {Turner},
  {Farrington}, \& {Goldfinger}}]{vonbraun2011}
{von Braun}, K., {Boyajian}, T.~S., {ten Brummelaar}, T.~A., {et~al.} 2011,
  \apj, 740, 49, \dodoi{10.1088/0004-637X/740/1/49}

\bibitem[{{Wilson}(1978)}]{Wilson1978}
{Wilson}, O.~C. 1978, \apj, 226, 379, \dodoi{10.1086/156618}

\bibitem[{{Winn} {et~al.}(2011){Winn}, {Matthews}, {Dawson}, {Fabrycky},
  {Holman}, {Kallinger}, {Kuschnig}, {Sasselov}, {Dragomir}, {Guenther},
  {Moffat}, {Rowe}, {Rucinski}, \& {Weiss}}]{Winn2011}
{Winn}, J.~N., {Matthews}, J.~M., {Dawson}, R.~I., {et~al.} 2011, \apjl, 737,
  L18, \dodoi{10.1088/2041-8205/737/1/L18}

\bibitem[{{Wisdom}(2005)}]{Wisdom2005}
{Wisdom}, J. 2005, in Bulletin of the American Astronomical Society, Vol.~37,
  AAS/Division of Dynamical Astronomy Meeting \#36, 525

\bibitem[{{Yi} {et~al.}(2004){Yi}, {Demarque}, \& {Kim}}]{Yi2004}
{Yi}, S.~K., {Demarque}, P., \& {Kim}, Y.-C. 2004, \apss, 291, 261,
  \dodoi{10.1023/B:ASTR.0000044330.92199.e2}

\end{thebibliography}

\end{document}